\documentclass[aps,superscriptaddress,preprintnumbers,amsmath,amssymb,a4paper,preprint]{revtex4}
\usepackage{graphicx}
\usepackage{dcolumn}
\usepackage{bm}
\usepackage{color}

\begin{document}

\title{Photoelectron circular dichroism of O 1$s$-photoelectrons of uniaxially oriented trifluoromethyloxirane: Energy dependence and sensitivity to molecular configuration}

\author{\firstname{G.} \surname{Nalin}}\email{nalin@atom.uni-frankfurt.de}
\affiliation{Institut f\"{u}r Kernphysik, Goethe-Universit\"{a}t, Max-von-Laue-Strasse 1, 60438 Frankfurt am Main, Germany}

\author{\firstname{K.} \surname{Fehre}}\email{fehre@atom.uni-frankfurt.de}
\affiliation{Institut f\"{u}r Kernphysik, Goethe-Universit\"{a}t, Max-von-Laue-Strasse 1, 60438 Frankfurt am Main, Germany}

\author{\firstname{F.} \surname{Trinter}}
\affiliation{Deutsches Elektronen-Synchrotron (DESY), Notkestrasse 85, 22607 Hamburg, Germany}
\affiliation{Molecular Physics, Fritz-Haber-Institut der Max-Planck-Gesellschaft, Faradayweg 4-6, 14195 Berlin, Germany}

\author{\firstname{N.~M.}\surname{ Novikovskiy}}
\affiliation{Institut f\"{u}r Physik und CINSaT, Universit\"{a}t Kassel, Heinrich-Plett-Strasse 40, 34132 Kassel, Germany}
\affiliation{Institute of Physics, Southern Federal University, 344090 Rostov-on-Don, Russia}

\author{\firstname{N.} \surname{Anders}}
\affiliation{Institut f\"{u}r Kernphysik, Goethe-Universit\"{a}t, Max-von-Laue-Strasse 1, 60438 Frankfurt am Main, Germany}

\author{\firstname{D.} \surname{Trabert}}
\affiliation{Institut f\"{u}r Kernphysik, Goethe-Universit\"{a}t, Max-von-Laue-Strasse 1, 60438 Frankfurt am Main, Germany}

\author{\firstname{S.} \surname{Grundmann}}
\affiliation{Institut f\"{u}r Kernphysik, Goethe-Universit\"{a}t, Max-von-Laue-Strasse 1, 60438 Frankfurt am Main, Germany}

\author{\firstname{M.} \surname{Kircher}}
\affiliation{Institut f\"{u}r Kernphysik, Goethe-Universit\"{a}t, Max-von-Laue-Strasse 1, 60438 Frankfurt am Main, Germany}

\author{\firstname{A.} \surname{Khan}}
\affiliation{Institut f\"{u}r Kernphysik, Goethe-Universit\"{a}t, Max-von-Laue-Strasse 1, 60438 Frankfurt am Main, Germany}

\author{\firstname{R.} \surname{Tomar}}
\affiliation{Institut f\"{u}r Kernphysik, Goethe-Universit\"{a}t, Max-von-Laue-Strasse 1, 60438 Frankfurt am Main, Germany}

\author{\firstname{M.} \surname{Hofmann}}
\affiliation{Institut f\"{u}r Kernphysik, Goethe-Universit\"{a}t, Max-von-Laue-Strasse 1, 60438 Frankfurt am Main, Germany}

\author{\firstname{M.} \surname{Waitz}}
\affiliation{Institut f\"{u}r Kernphysik, Goethe-Universit\"{a}t, Max-von-Laue-Strasse 1, 60438 Frankfurt am Main, Germany}

\author{\firstname{I.} \surname{Vela-Perez}}
\affiliation{Institut f\"{u}r Kernphysik, Goethe-Universit\"{a}t, Max-von-Laue-Strasse 1, 60438 Frankfurt am Main, Germany}

\author{\firstname{H.} \surname{ Fukuzawa}}
\affiliation{Institute of Multidisciplinary Research for Advanced Materials, Tohoku University, Sendai 980-8577, Japan}

\author{\firstname{K.} \surname{Ueda}}
\affiliation{Institute of Multidisciplinary Research for Advanced Materials, Tohoku University, Sendai 980-8577, Japan}

\author{\firstname{J.} \surname{ Williams}}
\affiliation{Department of Physics, University of Nevada, Reno, Nevada 89557, United States}

\author{\firstname{D.} \surname{Kargin}}
\affiliation{Institut f\"{u}r Chemie und CINSaT, Universit\"{a}t Kassel, Heinrich-Plett-Strasse 40, 34132 Kassel, Germany}

\author{\firstname{M.} \surname{Maurer}}
\affiliation{Institut f\"{u}r Chemie und CINSaT, Universit\"{a}t Kassel, Heinrich-Plett-Strasse 40, 34132 Kassel, Germany}

\author{\firstname{C.} \surname{K\"{u}stner-Wetekam}}
\affiliation{Institut f\"{u}r Physik und CINSaT, Universit\"{a}t Kassel, Heinrich-Plett-Strasse 40, 34132 Kassel, Germany}

\author{\firstname{L.} \surname{Marder}}
\affiliation{Institut f\"{u}r Physik und CINSaT, Universit\"{a}t Kassel, Heinrich-Plett-Strasse 40, 34132 Kassel, Germany}

\author{\firstname{J.} \surname{Viehmann}}
\affiliation{Institut f\"{u}r Physik und CINSaT, Universit\"{a}t Kassel, Heinrich-Plett-Strasse 40, 34132 Kassel, Germany}

\author{\firstname{A.} \surname{Knie}}
\affiliation{Institut f\"{u}r Physik und CINSaT, Universit\"{a}t Kassel, Heinrich-Plett-Strasse 40, 34132 Kassel, Germany}

\author{\firstname{T.} \surname{Jahnke}}
\affiliation{Institut f\"{u}r Kernphysik, Goethe-Universit\"{a}t, Max-von-Laue-Strasse 1, 60438 Frankfurt am Main, Germany}
\affiliation{European XFEL GmbH, Holzkoppel 4, 22869 Schenefeld, Germany}

\author{\firstname{M.} \surname{Ilchen}}\email{markus.ilchen@xfel.eu}
\affiliation{Deutsches Elektronen-Synchrotron (DESY), Notkestrasse 85, 22607 Hamburg, Germany}
\affiliation{Institut f\"{u}r Physik und CINSaT, Universit\"{a}t Kassel, Heinrich-Plett-Strasse 40, 34132 Kassel, Germany}
\affiliation{European XFEL GmbH, Holzkoppel 4, 22869 Schenefeld, Germany}

\author{\firstname{R.} \surname{D\"{o}rner}}
\affiliation{Institut f\"{u}r Kernphysik, Goethe-Universit\"{a}t, Max-von-Laue-Strasse 1, 60438 Frankfurt am Main, Germany}

\author{\firstname{R.} \surname{Pietschnig}}\email{pietschnig@uni-kassel.de}
\affiliation{Institut f\"{u}r Chemie und CINSaT, Universit\"{a}t Kassel, Heinrich-Plett-Strasse 40, 34132 Kassel, Germany}

\author{\firstname{Ph.~V.} \surname{Demekhin}}\email{demekhin@physik.uni-kassel.de}
\affiliation{Institut f\"{u}r Physik und CINSaT, Universit\"{a}t Kassel, Heinrich-Plett-Strasse 40, 34132 Kassel, Germany}

\author{\firstname{M.~S.} \surname{Sch\"{o}ffler}}\email{schoeffler@atom.uni-frankfurt.de}
\affiliation{Institut f\"{u}r Kernphysik, Goethe-Universit\"{a}t, Max-von-Laue-Strasse 1, 60438 Frankfurt am Main, Germany}

\date{\today}

\begin{abstract}
The photoelectron circular dichroism (PECD) of the O 1$s$-photoelectrons of trifluoromethyloxirane(TFMOx) is studied experimentally and theoretically for different photoelectron kinetic energies. The experiments were performed employing circularly polarized synchrotron radiation and coincidentelectron and fragment ion detection using Cold Target Recoil Ion Momentum Spectroscopy. The corresponding calculations were performed by means of the Single Center method within the relaxed-core Hartree-Fock approximation.  We concentrate on the energy dependence of the differential PECD of uniaxially oriented TFMOx molecules, which is accessible through the employed coincident detection.  We also compare results for differential PECD of TFMOx to those obtained for the equivalent fragmentation channel and similar photoelectron kinetic energy of methyloxirane (MOx), studied in our previous work. Thereby, we investigate the influence of the substitution of the methyl-group by the trifluoromethyl-group at the chiral center on the molecular chiral response. Finally, the presently obtained angular distribution parameters are compared to those available in literature.
\end{abstract}

%\pacs{33.80.-b, 32.80.Hd,  33.55.+b,  81.05.Xj}
%\keywords{Photon interactions with molecules; Auger effect and inner-shell excitation or ionization; Optical activity and dichroism; Chiral media}

\maketitle

\section{Introduction}
\label{sec:intro}

The interaction of circularly polarized light with chiral molecules leads to a range of chiroptical effects. One of such effects, namely the photoelectron circular dichroism (PECD), has been demonstrated to have strong potential as a sensitive probe of this chiral light-molecule interaction. The effect manifests itself as a forward-backward asymmetry of the photoelectron angular emission distribution with respect to the light propagation direction, which inverts upon switching of the light's helicity or upon exchanging molecular enantiomers. After the first theoretical formulation provided by Ritchie in 1976 \cite{Ritchie76} and the first experimental observation by B\"{o}wering \textit{et al.} in 2001 \cite{Bowering01}, PECD rapidly evolved into a well-established sensitive chiral recognition technique \cite{Powis08a,Nahon15,Turchini17,Hadidi18}.

PECD is a universal chiroptical effect occurring in all regimes \cite{Beaulieu16NJP}, from one-photon to multiphoton ionization  \cite{Lux12AngChm,Lehmann13jcp} and  strong-field ionization \cite{Beaulieu16NJP,Fehre19,Fehre19jpca}, with a majority of experiments, so far, reporting on the ionization of valence electrons \cite{Garcia03,Nahon06,Garcia08,Garcia10,Tia13}. PECD shows a complex dependence on the electronic \cite{Garcia14} and vibronic \cite{Garcia13} configuration, as well as the conformation \cite{Turchini13,Nahon16}, dimerization \cite{Nahon10}, clustering \cite{Daly11} of molecules,  and the energy of the emitted photoelectron \cite{Powis08b}. The typical strength of PECD is on the order of a few percent. Its unique sensitivity has been exploited to study ultrafast molecular dynamics \cite{Beaulieu16FD,Ilchen21tfmox}, especially time-resolved molecular relaxation \cite{Comby16}, and it has been employed for real-time determination of the enantiomeric excess  in racemic mixtures \cite{Comby18}.

Several experiments reported on PECD after core ionization of chiral molecules \cite{Stener04,Nahon06,Powis08b,Hergenhahn04,Alberti08,Ulrich08,Pitzer16cpc,Pitzer16jpb,Tia17,Ilchen17,Hartmann19}. In our  recent work \cite{Tia17},  we performed a differential study of PECD after O $1s$-photoionization of uniaxially oriented  methyloxirane molecules (MOx, C$_3$H$_6$O). To this end, the O $1s$-photoelectrons were detected in coincidence with two charged fragments. The molecular fragmentation axis was assumed to coincide with the emission direction of those  ionic fragments relying on the validity of the so-called  axial-recoil approximation \cite{Zare72}. This assumption is often well justified for cases of rapid Auger decay of an inner-shell vacancy  and  subsequent Coulomb explosion of a molecular dication. It was found that a selection of certain fragmentation directions of MOx with respect to the light propagation direction yields an increase of the PECD by a factor of 10, as compared to randomly oriented molecules.

\begin{figure}
\includegraphics[scale=.20]{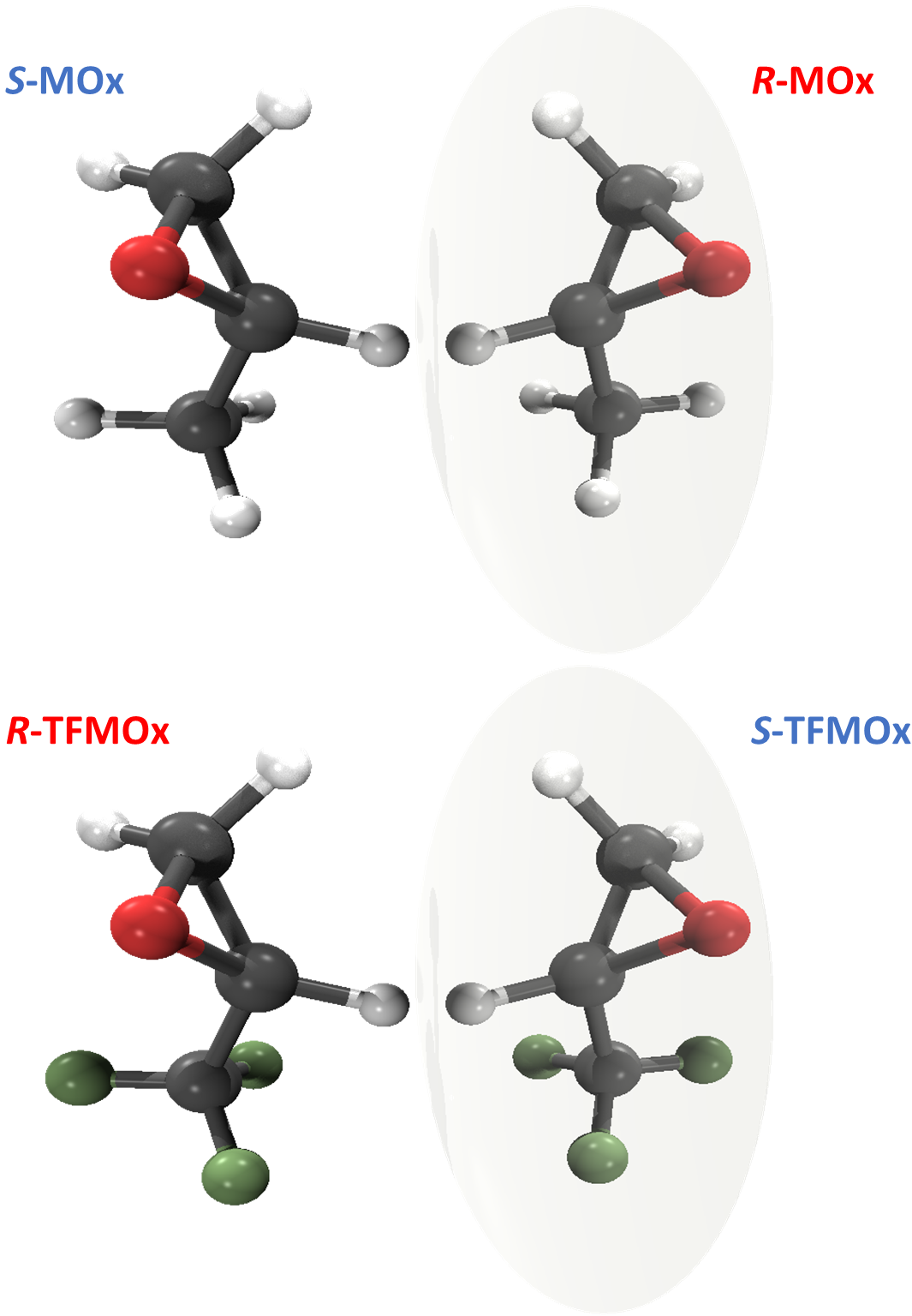}
\caption{Top: Structure of \textit{S-} and \textit{R-}methyloxirane (C$_3$H$_6$O, MOx). Bottom: Structure of \textit{R-} and \textit{S-}trifluoromethyloxirane (C$_3$H$_3$F$_3$O, TFMOx). Note that substitution of the methyl-group (CH$_3$) with the trifluoromethyl-group (CF$_3$) at the chiral center causes a renaming of the respective \textit{R-} and \textit{S-}enantiomers according to the CIP rule  \cite{Cahn66}.}\label{fig:Molcs}
\end{figure}

In the present work, we extend our previous study \cite{Tia17} to a close relative of MOx, the trifluoromethyloxirane (TFMOx, C$_3$H$_3$F$_3$O) molecule, in which all three hydrogen atoms of the methyl-group are replaced by much heavier fluorine atoms. For the further discussion, it is important to stress that by definition of the Cahn-Ingold-Prelog (CIP \cite{Cahn66}) rules, this substitution exchanges the assigned handedness (\textit{S-} to \textit{R-}enantiomer and vice versa) despite the unchanged connectivity of the molecular backbone and the stereo-center in MOx and TFMOx (see Fig.~\ref{fig:Molcs}). We investigate here the O $1s$-photoelectrons of TFMOx for different kinetic energies. One of these energies coincides with that of the study on  MOx presented in Ref.~\citenum{Tia17}. We selected, furthermore, for our investigation of TFMOx one of the fragmentation channels which was studied previously for MOx.

\begin{figure}
\centering
\includegraphics[width=0.60\textwidth]{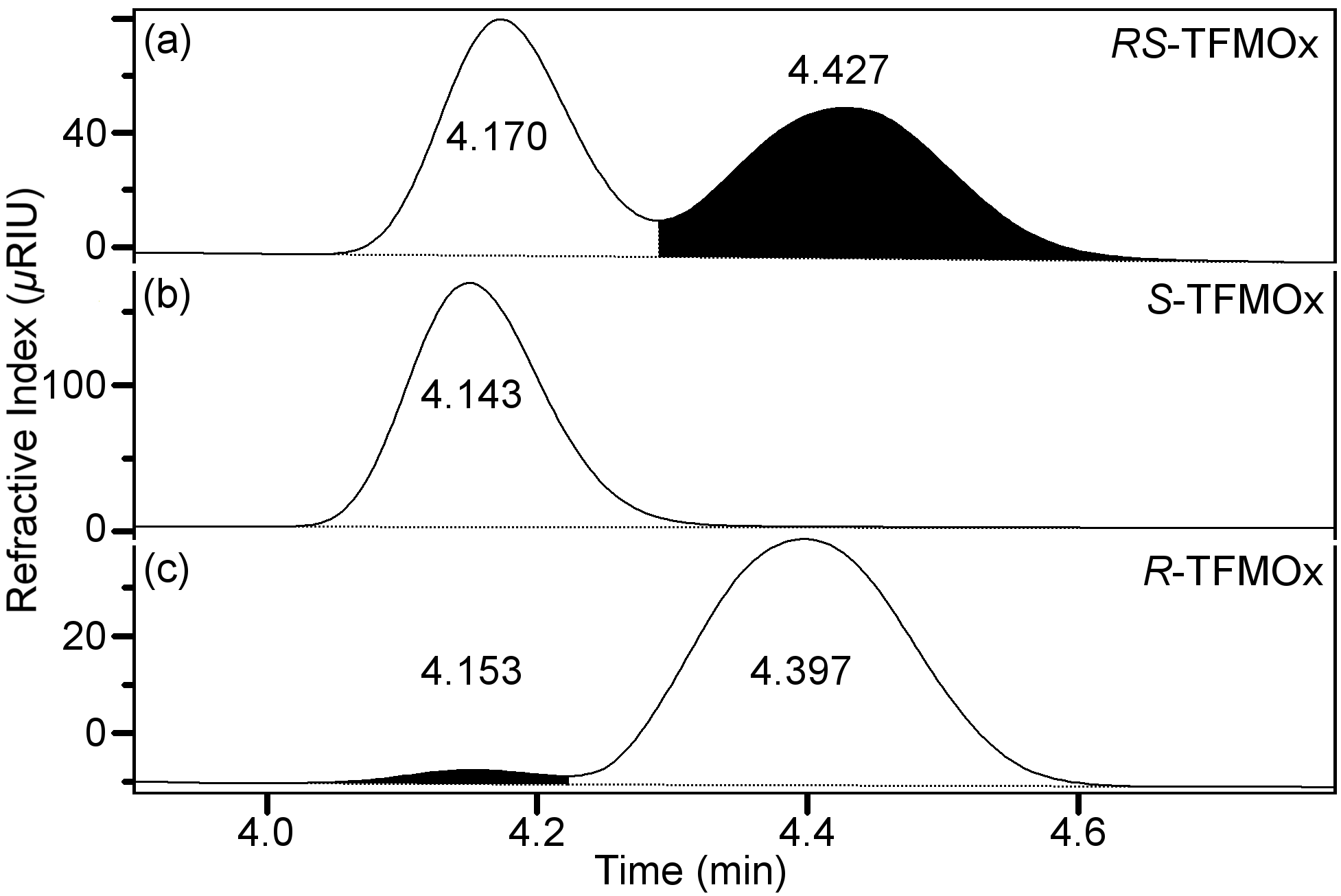}
\caption{Ratio of enantiomers present in the samples of racemic \textit{RS-}TFMOx (a) and its \textit{S-} and \textit{R-}enantiomers (b and c, respectively) as experimentally determined using HPLC.}\label{fig:EEpurity}
\end{figure}

\section{Research Methods}
\label{sec:Methods}

\subsection{Samples}
\label{sec:MethodsSamples}

An \textit{RS-}mixture and samples of pure \textit{S-} or \textit{R-}TFMOx enantiomers  were commercially obtained (SynQuest Laboratories) and used as received. Their chemical identity and purity were verified by nuclear magnetic resonance (NMR) spectroscopy. The  $^1$H, $^{13}$C, and $^{19}$F NMR data were recorded on Varian VNMRS-500~MHz or MR-400~MHz spectrometers at room temperature. Chemical shifts were referenced to residual protic impurities in the solvent ($^1$H) or the deuterio solvent itself ($^{13}$C) and reported relative to external Si(CH$_3$)$_4$ ($^1$H, $^{13}$C) or C$_6$H$_5$CF$_3$($^{19}$F).  The chemical purity after integration of the fluorine NMR signal has been estimated to be higher than 96\% for the \textit{RS-}TFMOx mixture and higher than 96\% or 98\% for the \textit{S-}TFMOx or \textit{R-}TFMOx enantiomers, respectively. The NMR shifts were assigned based on 2D-NMR (HSQC) spectra. The enantiomeric purity of the samples  was investigated by high-performance liquid chromatography (HPLC) using a CHIRALPAK IG (4.6~mm~$\times$~250~mm) chiral analytical column. According to these measurements, the \textit{RS-}TFMOx mixture is racemic within margins of error (S-TFMOx\,:\,\textit{R-}TFMOX = 49\%\,:\,51\%), and  the enantiopure \textit{S-}TFMOx or \textit{R-}TFMOx have an enantiomeric excess of approximately 95\% (see Fig.~\ref{fig:EEpurity} for details).

\subsection{Experiments}
\label{sec:MethodsExperiment}

The experiment was conducted during 8-bunch mode (pulsed operation) at  synchrotron SOLEIL  (Saint-Aubin, France) at beamline SEXTANTS using Cold Target Recoil Ion Momentum Spectroscopy (COLTRIMS, \cite{Doerner00,Ullrich03,Jahnke04}). The light's helicity was switched every two hours between circular left (CL) and circular right (CR). The samples were expanded through a nozzle of 60~$\mu$m diameter, and the expanding gas was skimmed two times, resulting in a free molecular gas jet of about 1~mm diameter. The jet was intersected at right angle with the synchrotron-radiation beam. A closed-loop recycling system was installed and used to reduce sample consumption \cite{Fehre20rsi}. After photoionization, the ionic fragments and electrons were accelerated by the electric field of the COLTRIMS spectrometer in opposite directions towards time- and position-sensitive detectors. A mesh-free spectrometer with electrostatic lenses and field-free drift regions was employed in order to improve the analyzer's angle and mass resolutions, and to increase the overall detection efficiency. The electric field at the point of interaction was set to $E=116~$V/cm. The time- and space-focusing geometry of the COLTRIMS spectrometer \cite{doerner1997nimb,Schoeffler2011njp, Schoeffler2013prl} resulted in an solid angle of acceptance of 4$\pi$ for both, photoelectrons up to 15~eV and ions up to 10~eV kinetic energy.

The electron energy calibration was carried out using a series of measurements on argon atoms, ionizing both, the $2p_{1/2}$ and $2p_{3/2}$ shells. From this series, a calibration function was created which links the measured electron positions-of-impact and times-of-flight with the photoelectron energy. The five photon energies for the TFMOx measurement (541.5, 542.5, 544.5, 546.5, and 550.1~eV) were chosen such that O $1s$-photoelectrons with kinetic energies of  3.1, 4.1, 6.1, 8.1, and 11.7~eV are created. The energy-calibration procedure covered this range of electron kinetic energies. The absolute and relative spatial orientation of the electron and ion detectors was checked comparing the molecular-frame photoelectron angular distributions of N$_2$ K-shell electrons measured during the same beamtime to those reported in Ref.~\citenum{Jahnke02}. The particles' impact positions and times-of-flight were measured by position- and time-sensitive microchannel plate (MCP) detectors, using delay-line anodes for position read-out (Roentdek GmbH). A HEX90 anode ($\varnothing$~90~mm) equipped with an 80~mm diameter MCP \cite{Jagutzki02} was used for the electron, and a HEX125L anode ($\varnothing$~125~mm) equipped with an efficiency-enhanced 75~mm diameter funnel MCP (Hamamatsu) \cite{Fehre18} for the ion detection. The momentum vector of each fragment at the instant of ionization is derived from their time-of-flight and impact position on the detector. Data were recorded with typical rates of 6~kHz on the ion and 10~kHz on the electron detector. The combined CF$_3^+$($m/z=69$)\,--\,C$_2$H$_{i=1,2,3}$O$^+$($m/z=41,42,43$) breakup channel contains roughly $2.5\times10^5$ valid events for each enantiomer, light helicity, and photon energy.

In order to cross-check the experimental results presented here,  we performed a second short experiment at the permanent COLTRIMS end-station of the PIPE instrument, located at beamline P04 of the synchrotron PETRA~III (DESY, Hamburg, Germany) \cite{Viefhaus13}. In this case, hexagonal delay-line detectors with an active diameter of 80~mm (HEX90 anodes) were used for both, ion and electron detection. The ion arm of that spectrometer was a 17~cm long acceleration region without any meshes, while the electron arm was built in a time-focussing geometry \cite{Wiley1955rsi}. The dichroic parameter $\beta_1$ was retrieved from the measured data by integrating over all possible fragmentation channels of TFMOx. Therefore, it was sufficient to record datasets for each polarization direction and photon energy for only 20 minutes. The data were recorded at detection rates of 15~kHz for the electron detector. The photoelectron-energy calibrations as well as detector-orientation calibrations were similarly performed as described above for the measurements at  synchrotron SOLEIL.

\begin{figure}
\centering
\includegraphics[width=0.60\textwidth]{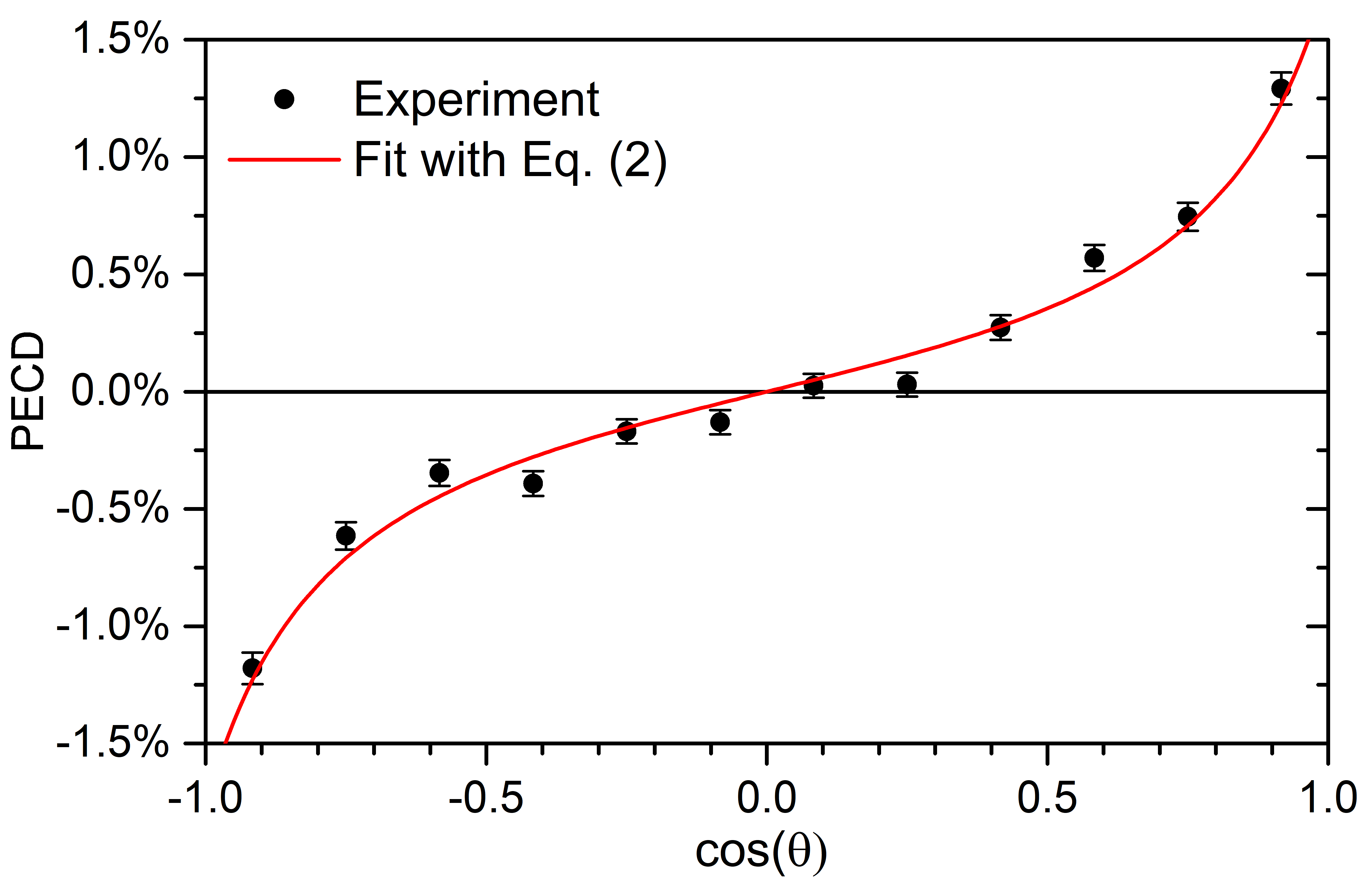}
\caption{Retrieval of the anisotropy and dichroic parameters from the measured PECD of \textit{R-}TFMOx recorded at the O 1s-photoelectron energy of 11.7~eV. A fit (red line) of the experimental data (circles with error bars)  using Eq.~(\ref{eq:PECD}) yielded $\beta_1=0.76\%\pm0.06\%$ and $\beta_2=1.14\pm0.15$.}\label{fig:Analysis}
\end{figure}

\subsection{Data analysis}
\label{sec:MethodsData}

Within the electric-dipole approximation, the laboratory-frame photoelectron angular emission distribution can be expanded in terms of Legendre polynomials $P_L$ via the dichroic parameter $\beta_1$ and the anisotropy parameter $\beta_2$. For randomly oriented molecules and in case of an ionization employing circularly polarized photons (with helicity $\pm 1$), the corresponding expansion reads:
\begin{equation}
I_{\pm1}(\theta) \propto  1 \pm \beta_1 P_1(\cos \theta) -\frac{1}{2} \beta_2 P_2(\cos \theta), \label{eq:DPICS}
\end{equation}
where the emission angle $\theta$ is defined with respect to the light propagation direction \cite{Cherepkov82}. We utilize one of the standard definitions of PECD \cite{Hergenhahn04,Tia17}, which is given by the normalized difference of the angular distributions (\ref{eq:DPICS}) recorded for left- and right-handed circularly polarized light (referred to as $+1$ and $-1$, respectively):
\begin{equation}
PECD(\theta)=\frac{I_{+1}(\theta)-I_{-1}(\theta)}{I_{+1}(\theta)+I_{-1}(\theta)}=\frac{\beta_1 P_1(\cos \theta)}{ 1-\frac{1}{2} \beta_2 P_2(\cos \theta)}. \label{eq:PECD}
\end{equation}
We used this equation to extract the parameters $\beta_1$ and  $\beta_2$ from the  measured $\theta$-dependent PECD. Figure~\ref{fig:Analysis} illustrates the fitting procedure for a photoelectron kinetic energy of 11.7~eV using \textit{R-}TFMOx enantiomers as a target. In this example, the fitting procedure yields $\beta_1=0.76\%\pm0.06\%$ and $\beta_2=1.14\pm0.15$. The uncertainties represent the standard error of the  fitting.

In order to check for possible systematic experimental errors, various subsets of the experimental data were considered separately for this analysis. For example, the high electric fields of the spectrometer degraded the time-of-flight resolution and thus the resolution of the momentum component parallel to the spectrometer axis. As one cross-check, only the central part of the photoelectron time-of-flight distribution was examined and analyzed as a subset. Furthermore, in order to exclude possible influences of  inhomogeneous detection probabilities, the electron detector was divided into its left and right half along the light propagation direction, and each subset was analyzed separately. The influence of possible ion feedback from the ion detector was investigated with the help of applying different conditions on the ion time-of-flight and by a selection of different ranges of sums of momenta for various fragmentation channels. All these subsets of the data yielded (within statistical errors) the same angular distribution  parameters $\beta_1$ and $\beta_2$.

\subsection{Theory}
\label{sec:MethodsTheory}

The O $1s$-photoionization of TFMOx was described  by the \textit{ab initio} theoretical approach used in our previous angle-resolved studies on MOx \cite{Tia17,Hartmann19} and TFMOx \cite{Ilchen17} molecules.  The electronic structure and dynamics calculations were carried out by employing the Single Center (SC)  method and code \cite{Demekhin11,Galitskiy15}, which provide an accurate description of the partial angular momentum photoelectron continuum waves of ionized molecules. The calculations were performed for the equilibrium molecular geometry of the neutral ground electronic state of TFMOx. Details of the calculations can be found in our previous study of TFMOx \cite{Ilchen17}. Therefore, only essential differences between the previous and present calculations are discussed below. While the calculations in  Ref.~\citenum{Ilchen17} were carried out within the frozen-core Hartree-Fock (FCHF) approximation, in the present work, the photoionization transition amplitudes were computed within the relaxed-core Hartree-Fock (RCHF) approximation, which accounts for the major effect of the monopole relaxation of molecular orbitals induced by the creation of the inner-shell vacancy. In addition, we increased here the SC expansions of the occupied orbitals by partial harmonics with $\ell_c,\vert m_c\vert \leq 79$ (as compared to $\ell_c,\vert m_c\vert \leq 59$ in Ref.~\citenum{Ilchen17}), while those for the partial photoelectron continuum waves were kept unchanged with  $\ell_\varepsilon,\vert m_\varepsilon\vert \leq 29$.

The calculations were performed for different orientations of the molecular frame (MF) with respect to the laboratory frame (LF, $Z_{LF}$ is defined by the propagation direction of the circularly polarized light), which is given by the two Euler angles $\alpha$ and $\beta$. Note, that the third angle $\gamma$ (which defines a rotation around the laboratory $Z_{LF}$ axis) is irrelevant in case of using circularly polarized light.  In the experiment, a breakup into two molecular fragments was examined. Thus, only a single fragmentation axis was determined. Therefore, the corresponding differential $PECD(\theta,\beta)$  was obtained by numerical integration of the computational results over the orientation angle $\alpha$. Here, the orientation interval $\alpha \in [0,2\pi)$  was covered in steps of $\Delta\alpha=0.025\pi$. Since the fragmentation dynamics can be rather complicated (and not all of the particles are detected in coincidence), it is not straightforward to relate the fragmentation axis obtained in the experiment from the asymptotic momenta of the two detected ions to a given molecular orientation at the instant of ionization. Therefore, just as in our previous work on  uniaxially oriented MOx \cite{Tia17}, we used the molecular orientation axis at the instant of the photoionization process as a free parameter in the calculations, and searched for the best visual correspondence between the computed and measured differential PECD. Details on this procedure can be found in the supplemental material document of our previous work \cite{Tia17}.

\section{Results and discussion}
\label{sec:Results}

\subsection{Differential PECDs: MOx vs. TFMOx}
\label{sec:ResultsPECDs}

\begin{figure}
\centering
\includegraphics[width=0.65\textwidth]{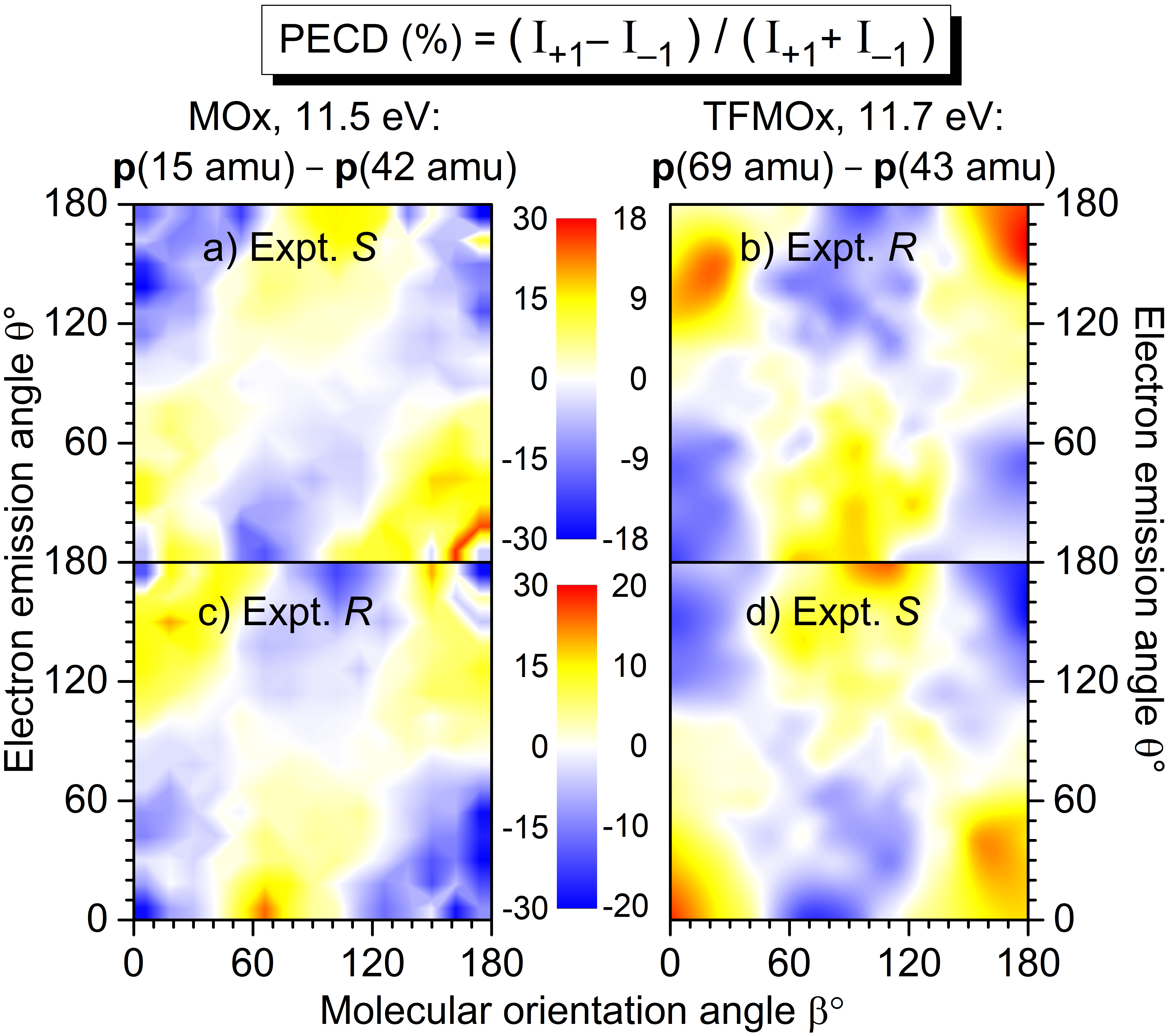}
\caption{Measured differential PECD of the uniaxially oriented enantiomers of MOx (left column, data from Ref.~\citenum{Tia17}) and TFMOx (right column, present work) as a function of the photoelectron emission angle $\theta$ and the molecular orientation angle $\beta$. The PECD maps were obtained for O $1s$-photoionization yielding similar electron kinetic energies (11.5~eV for MOx and 11.7~eV for TFMOx). Conceptually the same fragmentation channels are chosen, i.e., CH$_3^+$($m/z=15$)\,--\,C$_2$H$_{2}$O$^+$($m/z=42$) for MOx and CF$_3^+$($m/z=69$)\,--\, C$_2$H$_{i=1,2,3}$O$^+$($m/z=41,42,43$) for TFMOx. According to the CIP rule, opposite enantiomers of these two chiral molecules with identical geometrical structure are compared in each row  (see Fig.~\ref{fig:Molcs} for details).} \label{fig:compare}
\end{figure}

In our previous work on uniaxially oriented MOx \cite{Tia17}, a differential PECD was studied for a photoelectron kinetic energy of 11.5~eV and two fragmentation channels. One of the channels represents a single-bond breakage with the loss of the methyl-group, i.e., a breakup into CH$_3^+$($m/z=15$)\,--\,C$_2$H$_{2}$O$^+$($m/z=42$). Therefore, we decided to focus in the present work on a similar fragmentation channel of TFMOx, which represents a single-bond breaking with the loss of the trifluoromethyl-group. We observed that additional losses of hydrogen atoms on the oxirane ring do not alter the measured electron emission distributions. Therefore, we examined here the differential PECD combining the datasets of the breakup channels  CF$_3^+$($m/z=69$)\,--\,C$_2$H$_{i=1,2,3}$O$^+$($m/z=41,42,43$), which we designated as   $\mathbf{p}(69~\mathrm{amu})-\mathbf{p}(43~\mathrm{amu})$ for brevity. The relative momentum (i.e., the difference of the momentum vectors) of those two fragments defines a fragmentation axis of the molecule, and its orientation Euler angle $\beta$ in the laboratory frame (with respect to the light propagation direction) is thus known. The molecule is that way, however, only uniaxially oriented which leaves one degree of freedom for its rotation around this axis.

We first consider the photoelectron kinetic energy of 11.7~eV, which is very similar to that used in Ref.~\citenum{Tia17} for MOx. The measured differential PECDs of uniaxially oriented enantiomers of TFMOx are depicted in the right column of Fig.~\ref{fig:compare} in panels (b) and (d). For some molecular orientations (i.e., some spatial orientations of the molecular fragmentation), the differential PECD reaches about 20\%. Our measurements thus confirm the conclusion of Ref.~\citenum{Tia17} that fixing already one axis of a chiral molecule in space enhances the chiral asymmetry by about an order of magnitude (as compared to about 1\% observed for the randomly oriented molecules shown in Fig.~\ref{fig:Analysis}). The asymmetry switches its sign upon exchanging of the enantiomers, which confirms the chiral origin of the observed effect. This normalized difference obeys also the analytically derived asymmetry property \cite{Tia17}: $PECD(\pi-\theta,\pi-\beta)=-PECD(\theta,\beta)$, which implies a flip of its sign upon rotation of this two-dimensional map by $180^\circ$ around the normal to the picture plane. Please note that this equivalence of changing the sign of the light helicity and  exchanging the enantiomer holds only for enantiomeric pure samples and perfectly circularly polarized light (see Ref.~\citenum{Fehre19} for more details).

For comparison, panels (a) and (c) in the left column of  Fig.~\ref{fig:compare} show the differential PECD of MOx from Ref.~\citenum{Tia17}.  One can  see that---for equal fragmentation channels and similar photoelectron kinetic energies---the differential PECDs  observed for these two molecules are very similar in the strengths and structures. It is surprising that exchanging the three hydrogen atoms in the methyl-group by the much heavier scatterers (fluorine atoms) does not significantly change the observed PECD landscape. In each row of this figure, we compare the data for MOx and TFMOx molecules of similar geometrical structures (therefore, opposite \textit{R-} and \textit{S-}enantiomers). One can see that the differential PECDs of the two molecules of similar geometrical structures have opposite-sign landscapes, whereas for identical  \textit{R-} or \textit{S-}enantiomers of both molecules the measured sign landscapes are very similar. This fact confirms that the inner-shell PECD effect emerges as a result of a complex multiple scattering of the emitted photoelectron wave by the molecular potential. Indeed, similar stereo-descriptors of enantiomers assume a similar circular arrangement of the stronger-to-weaker scatterers,  which couple rotational motion of a photoelectron, driven by the electric field vector of circularly polarized light, with its translational motion  (like a nut on a bolt \cite{Powis08b,Tia17}).

\subsection{Energy dependence of PECD}
\label{sec:ResultsEnergys}

\begin{figure}
\centering
\includegraphics[width=0.65\textwidth]{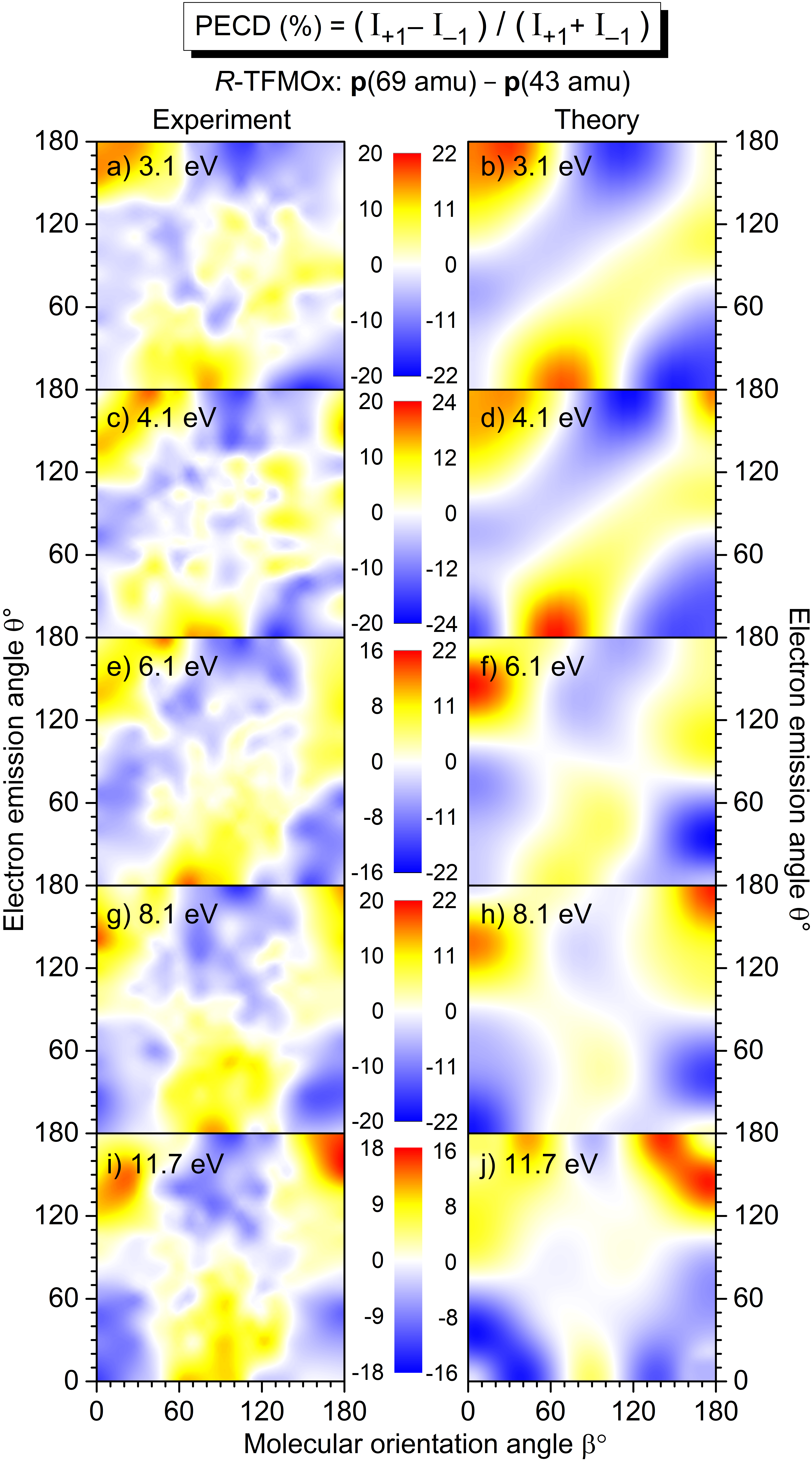}
\caption{Experimental (left column) and theoretical (right column) differential PECD of  uniaxially oriented \textit{R}-TFMOx as a function of the photoelectron emission angle $\theta$ and the molecular orientation angle $\beta$, obtained for different kinetic energies (indicated in each panel) of the O $1s$-photoelectron and the combined  fragmentation channels CF$_3^+$($m/z=69$)\,--\, C$_2$H$_{i=1,2,3}$O$^+$($m/z=41,42,43$).} \label{fig:energy}
\end{figure}

A full account of the differential PECD of uniaxially oriented TFMOx, measured for the five aforementioned photoelectron kinetic energies, is depicted in the left column of Fig.~\ref{fig:energy}. For all energies, there are molecular orientations at which the measured differential PECD reaches maximal values of about $\pm 16$ to $\pm 20\%$ [panels (a), (c), (e), (g), and (i) in the left column], which are at least larger by a factor of 10 than the respective orientation-averaged values $\beta_1$ (see Sec.~\ref{sec:ResultsParameters} and Fig.~\ref{fig:ADparam}). Qualitatively, all measured differential PECDs depict  similar landscapes including their signs. Nevertheless, a clear trend in the development of the measured differential PECDs with the increase of  the photoelectron kinetic energy can be noticed. For the two lower kinetic energies of 3.1~eV in panel (a) and  4.1~eV in panel (c), as well as 6.1~eV in panel (e), we observe a sequence of about 45$^\circ$-inclined positive-negative-positive-negative stripes (from top to bottom). For the two higher kinetic energies of 8.1~eV in panel (g) and 11.7~eV in panel (i), it develops in a clear sequence of three horizontal positive-negative-positive islands at the top and an alternated sequence at the bottom.

In order to calculate the differential PECD of uniaxially oriented TFMOx, we first assumed validity of the axial-recoil approximation \cite{Zare72}, for which the momentum difference of two fragments represents an orientation of the molecular axis in space at the  instant of ionization (i.e., the molecule does not rotate significantly during the Auger decay and subsequent fragmentation). This approximation worked well in our previous study \cite{Tia17}. There, we found that---for the single-bond breaking with the loss of the methyl-group---the \textit{optimal}  molecular axis, which provides the best description of the experiment, connects the oxygen atom with the carbon of the methyl-group (see supplemental material document of Ref.~\citenum{Tia17}). Starting with this assumption, we searched for the \textit{optimal} molecular fragmentation axis, which provides the best match between theory and experiment for the  differential PECDs of  uniaxially oriented TFMOx  for all five electron energies.

As a result of our search, we find that the   best visual correspondence between the  measured and computed differential PECD is provided by a fragmentation axis which connects the oxygen atom with the CH$_2$-group on the oxirane ring.  It should be stressed that all five photoelectron energies yielded  very similar  \textit{optimal} fragmentation axes which coincide  within a solid angle of about 0.2~sr. The fact, that the axis obtained this way does not connect two charged fragments (the trifluoromethyl-group and the oxirane ring) but lies almost in the plane of the oxirane ring, indicates a breakdown of the axial-recoil approximation for the TFMOx molecule. Indeed, a considerably larger mass of the trifluoromethyl-group (as compared to the methyl-group in MOx) and a much richer electronic structure of fluorine atoms (as compared to hydrogens) may lead to the fact, that the centers of charge of the fragments do not coincide with their centers of mass. In this case, an intricate rotation of two fragments with respect to each other takes place in the course of a considerably slower dissociation of TFMOx (as compared to the same fragmentation channel of MOx with the much lighter methyl-group), and, as a consequence, the differences of their asymptotic momenta do not represent the molecular orientation at the instant of photoionization.

The results of our calculations of the  differential PECD of \textit{R-}TFMOx, obtained for the \textit{optimal} fragmentation axis, are depicted in the right column of Fig.~\ref{fig:energy}.  The calculations provide a semi-quantitative theoretical description of the measured differential PECD. Indeed, the overall landscapes and signs of the experimental differential PECDs, including the arrangement of stripes and islands discussed above,  are reproduced by the theory. Detailed distributions of the experimental and theoretical asymmetries in their forms and strengths are, however, somewhat different. This leads to the fact that averaging the theoretical and experimental signals over molecular orientations yields different $\beta_1$ values. This fact can be seen, e.g., for the photoelectron energies of 8.1 and 11.7~eV in   the forward emission direction. In particular in the experiment,  molecular orientations yielding positive chiral asymmetries dominate over those which yield a negative PECD, while this situation is opposite in the theory [cf., in Fig.~\ref{fig:energy} panels (g) and (h), and  separately panels  (i) and (j), for the emission angles around  $\theta=0^\circ$ and orientation angles  around $\beta=90^\circ$ with highest contribution weights]. This difference is reflected in opposite signs of the computed and measured dichroic parameters at 8.1 and 11.7~eV [see panel (a) of Fig.~\ref{fig:ADparam} and its discussion in Sec.~\ref{sec:ResultsParameters}].

\subsection{Anisotropy and dichroic parameters}
\label{sec:ResultsParameters}

\begin{figure}
\centering
\includegraphics[width=0.65\textwidth]{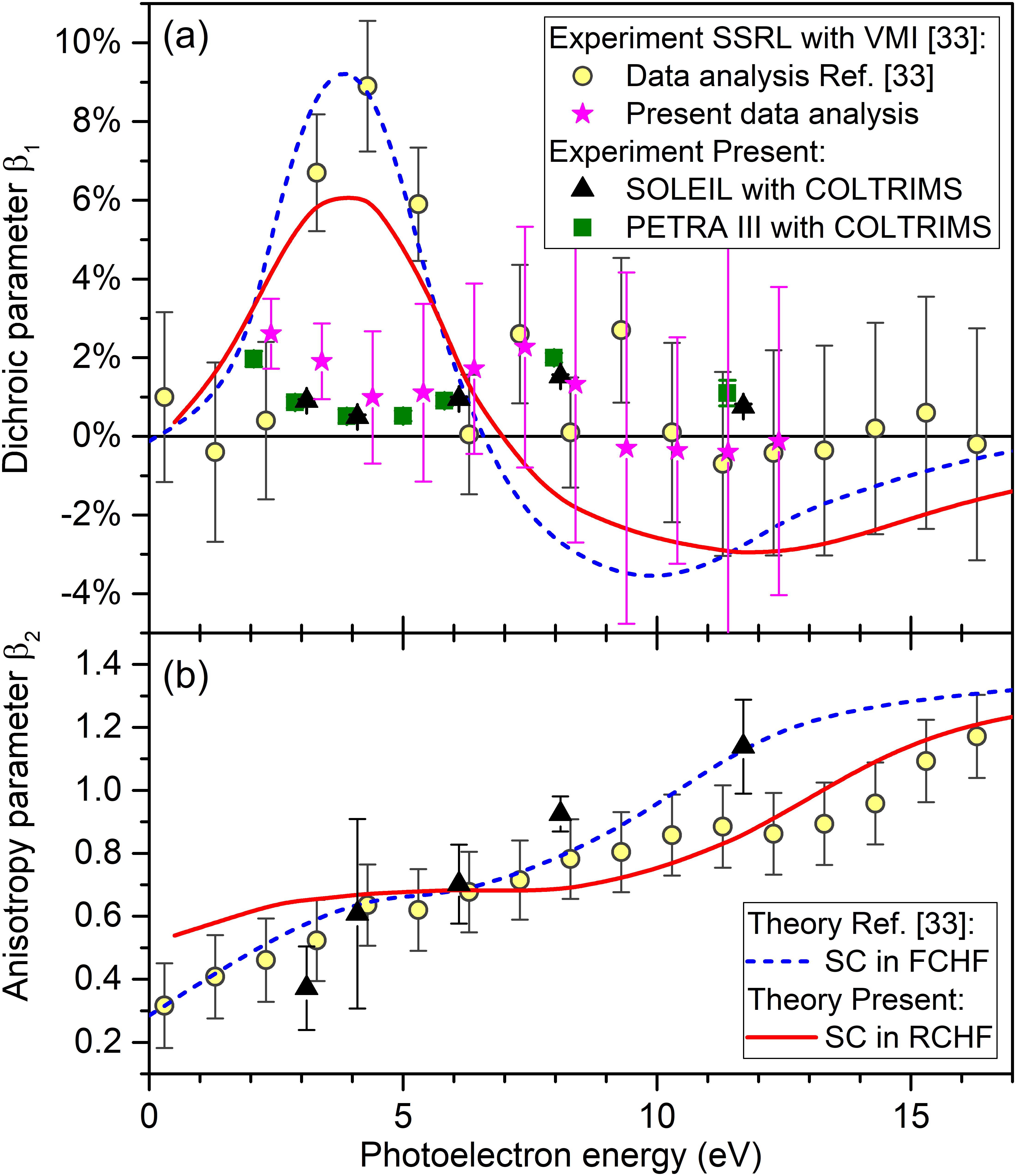}
\caption{Dichroic parameter $\beta_1$ (a)  and the anisotropy parameter $\beta_2$ (b) as given by Eq.~(\ref{eq:DPICS}) for the O $1s$-photoelectrons of \textit{R-}TFMOx. Circles: experimental results recorded at the BL13-2 beamline of SSRL using a VMI setup, as reported in Ref.~\citenum{Ilchen17}. Stars: $\beta_1$ obtained by the present reanalysis of the raw data underlying the values shown by circles from Ref.~\citenum{Ilchen17} (see Sec.~\ref{sec:ResultsParameters}). Triangles: results of the present experiment performed at beamline SEXTANTS of SOLEIL using COLTRIMS. Squares: corresponding  $\beta_1$ values measured at beamline P04 of PETRA III. Dashed curve: frozen-core Hartree-Fock calculations performed with the SC method as reported in Ref.~\citenum{Ilchen17}. Solid curve: present relaxed-core Hartree-Fock calculations performed with the SC method.}\label{fig:ADparam}
\end{figure}

The  angular distribution parameters $\beta_1$ and $\beta_2$ of the O $1s$-photoelectrons of TFMOx, measured at SOLEIL and defined by Eq.~(\ref{eq:DPICS}), are depicted in Fig.~\ref{fig:ADparam}  by black triangles. Panel (b) reveals that our experimental results for the anisotropy parameter $\beta_2$ are in a good agreement with the experimental and theoretical results reported in Ref.~\citenum{Ilchen17}. The latter were computed by using the SC method within the FCHF approximation and are shown as a blue dashed curve. Contrary to that, however, our measured results for the dichroic parameter $\beta_1$ disagree considerably with the experimental and theoretical results reported in Ref.~\citenum{Ilchen17} in an energy range between 3 and 6~eV. As a cross-check, the dichroic parameter $\beta_1$ obtained from a second experiment (see Sec.~\ref{sec:MethodsExperiment} for details) with a different setup at a different synchrotron-radiation facility is shown in  panel~(a) of Fig.~\ref{fig:ADparam} by green squares. It confirms the results from our main experiment (black triangles).

In order to understand the considerable deviation between the presently and previously measured dichroic parameters $\beta_1$ seen in panel~(a) of Fig.~\ref{fig:ADparam} in the energy range of 3--6~eV, in a joint effort with the authors of Ref.~\citenum{Ilchen17}, we reanalyzed the original Stanford Synchrotron Radiation Lightsource (SSRL) raw data recorded at the BL13-2 beamline with a VMI setup. The reanalysis of raw images was performed independently for every run of a certain photon energy, helicity of the light, and enantiomer, using the following steps. The intensity in the surrounding of the momentum spheres was subtracted as background. Then, the momentum sphere had to be centered. It turned out that the center varied from run to run on the level of a few percent of the radius of the momentum sphere giving rise for systematic errors if not corrected for. This shift had not been accurately corrected for in the original analysis in Ref.~\citenum{Ilchen17}. An Abel transformation was applied to the centered distributions followed by a transformation into polar coordinates. Finally only those electrons, which have the proper kinetic energy, are selected for further analysis. This is done by gating on a certain thickness of the momentum sphere of the Abel-inversed data. The resulting subset of data (intensity as function of the emission angle with respect to the light propagation axis) is treated analogue as the COLTRIMS data (see Sec.~\ref{sec:MethodsData}). Results of this reanalysis of the SSRL data from Ref.~\citenum{Ilchen17} are shown in panel~(a) of Fig.~\ref{fig:ADparam} by pink stars (note that the reanalysis was performed only in the relevant energy range of the present experiment). The specified uncertainties correspond to the fit error from the extraction of the $\beta_1$ parameter. A subdivision of each run from the SSRL measurement into two independent data sets confirmed both, the retrieved  values of the $\beta_1$ parameter and the respective uncertainties. As one can see from this figure, the dichroic parameters determined in all three experiments are now consistent.

In order to find a possible reason for the disagreement between the previously computed and all measured dichroic parameters in panel (a) of Fig.~\ref{fig:ADparam}, we cross-checked the computational results of Ref.~\citenum{Ilchen17}. First of all, we reexamined the equilibrium geometry of the molecule obtained in Ref.~\citenum{Ilchen17} on the (2,2)-CASSCF/6-31+G(d,p) level of theory and used in the subsequent calculations. For this purpose, we now performed its optimization with DFT methods using the B3LYP hybrid functional with a 6-311+G(d,p) basis set. The minimal-energy structure was confirmed by the absence of imaginary frequencies. It turned out that both equilibrium structures differ only negligibly and the angular distribution parameters recomputed here in the FCHF approximation basically confirm those computational results of Ref.~\citenum{Ilchen17} (not shown here for brevity). As a next step, we recomputed the angular distribution parameters of the O $1s$-photoelectrons of TFMOx using the RCHF approximation. It is known to be more accurate for core ionization than the FCHF approximation used in the previous work \cite{Ilchen17}.  The results of these calculations are shown in  Fig.~\ref{fig:ADparam} by red solid curves. As one can see, the effect of the core relaxation caused by the created inner-shell vacancy does not significantly change the energy dependence of the computed angular distribution parameters, and the presently  computed  $\beta_1$ and $\beta_2$  parameters agree well with the previously computed and measured data from Ref.~\citenum{Ilchen17} (blue dashed curve and yellow circles). For the dichroic parameter $\beta_1$, which is of primary interest here, relaxation of the core has just somewhat reduced its maximal and minimal values. A reason for the remaining disagreement between the theoretical and experimental dichroic parameters remains unknown (see also discussion in Sec.~\ref{sec:conclud}).

\section{Conclusions and outlook}
\label{sec:conclud}

A joint experimental and theoretical study of the photoelectron circular dichroism (PECD) of the  O $1s$-photoelectrons of uniaxially oriented  trifluoromethyloxirane (TFMOx) has been presented. We report the differential PECD for the combined single-bond breakup channels  CF$_3^+$($m/z=69$)\,--\,C$_2$H$_{i=1,2,3}$O$^+$($m/z=41,42,43$) with the loss of the trifluoromethyl-group at five photoelectron kinetic energies. The differential PECD of TFMOx, measured for a photoelectron kinetic energy of 11.7~eV, is very similar to that observed in our previous work \cite{Tia17} for an almost equal (11.5~eV)  photoelectron kinetic energy of the O $1s$-photoelectrons and the (conceptually) same fragmentation channel (single-bond breaking with the loss of the methyl-group) of methyloxirane (MOx). The measured PECD landscapes exhibit equal signs if one compares identical \textit{S-} or \textit{R-}enantiomers of those two molecules which possess similar hierarchic stereo arrangement of scatterers, and opposite signs for different enantiomers which have similar geometrical structures.

The molecular fragmentation axis, found theoretically for this breakup channel, does not connect the CF$_3^+$ and C$_2$H$_{3}$O$^+$ fragments, which is a clear indication that for larger molecules, the connection between the orientation of the molecule at the instant of photoionization and the asymptotic momenta of the detected fragments have a rather intricate interconnection. The computed differential PECDs reproduce those experimental results at all photoelectron kinetic energies on a semi-quantitative level. Having  rather similar landscapes, the computed and measured differential PECDs differ  in  their forms and maximal asymmetry strengths.  As a consequence, averaging of those rather strong asymmetries over all molecular orientations yields different small values of the theoretical and experimental dichroic parameters $\beta_1$. We performed a reanalysis of the raw data from the experiment published in Ref.~\citenum{Ilchen17} leading to a revision of some of the data points previously reported. These corrected previous experimental results agree with new results from two different experiments performed at two light sources in the present work.

Being very similar in their geometrical structures, \textit{S/R-}MOx and \textit{R/S-}TFMOx molecules are very different in their chemical and, thus, electronic structures. This could be a reason for a very different degree of agreement between theory and experiment in our earlier study of MOx \cite{Tia17} and in the present and earlier  studies of TFMOx \cite{Ilchen17}. Because of considerably more electrons in TFMOx, the respective calculations seem to be much more sensitive to minor details of the molecular potential, especially for slow photoelectrons. Experimentally, very different masses of constituent parts in TFMOx make an unambiguous assignment of a molecular orientation axis to a certain fragmentation channel impossible. In order to clarify the remaining disagreement between the theoretical and experimental dichroic parameters $\beta_1$, the O $1s$-photoionization of TFMOx needs to be reinvestigated with alternative theoretical methods (e.g.,  TDDFT \cite{Stener05} or CMS-X$\alpha$ \cite{Powis00}). Reference~\citenum{Stener04} reports a comprehensive theoretical study of the photoelectron angular distributions from chiral derivatives of oxirane by the former theoretical approach (note that TFMOx was not studied there). As one can learn from Fig.~8 of Ref.~\citenum{Stener04}, dichroic parameters computed for  O 1s-photoelectrons of all considered oxirane derivatives exhibit a clear enhancement to about 5--8\% within the photoelectron kinetic energy interval of 2--8~eV, supporting our theoretical findings on TFMOx. Taking also into account the robustness of the present and previous \cite{Ilchen17} calculations performed by the SC method, a reason for the remaining disagreement between the theoretical and experimental dichroic parameters $\beta_1$ can lie beyond the (fixed-nuclei  electric-dipole approximation) electronic structure calculations performed here.

\section*{Authors contribution}
The experiment was prepared and carried out by G.N., K.F., F.T.,N.A., D.T., S.G., M.K., A.K., R.T., M.H., M.W., I.V.-P., G.K., J.S.,D.T., H.F., K.U., J.B.W., C.K.-W., L.M., J.V., A.K., T.J., and M.S.S.Samples were analyzed by D.K., M.M., and R.P. Data analysis wasperformed by G.N., K.F., M.I., and M.S.S. Theoretical calculationswere  performed  by  N.M.N.  and  P.V.D.  Initial  draft  was  createdby G.N., K.F., R.D., P.V.D., and M.S.S. All authors discussed theresults and commented on the manuscript.

\section*{Conflicts of interest}
There are no conflicts to declare.

\section*{Acknowledgements}
This work was funded by the Deutsche Forschungsgemeinschaft (DFG)--Project No.\,328961117--SFB 1319 ELCH (Extreme light for sensing and driving molecular chirality). This research was also undertaken as part of the ASPIRE Innovative Training Network, which has received funding from the European Union's Horizon 2020 research and innovation programme under the Marie Sklodowska-Curie Grant Agreement No. 674960. H.F. and K.U. acknowledge the XFEL Priority Strategy Program of MEXT, the Research Program of `Dynamic Alliance for Open Innovations Bridging Human, Environment and Materials' and IMRAM project for support. K.F. acknowledges support by the German National Merit Foundation. M.I. acknowledges funding of a Peter-Paul-Ewald Fellowship by the Volkswagen foundation.  M.S.S. thanks the Adolf-Messer Foundation for financial support. We thank the staff of SOLEIL for running the facility and providing beamtime under project 20180746 and beamline SEXTANTS for excellent support. We also thank the staff of PETRA III for running the facility and providing beamtime under project  H-20010092 and beamline P04 for excellent support. The work at PETRA III was supported by BMBF.

\end{document}